\documentstyle{article}

\def\i{\wedge^b}
\def\ili{\vee^b}
\def\ol{\overline}
\def\di{\odot}
\def\dili{\oplus}

\title{On Boolean reliability algebra}
\author{Branislav Bori\v{c}i\'c, \tt boricic@ekof.bg.ac.rs\\ Mirjana Ili\'c, \tt mirjanailic@ekof.bg.ac.rs\\ Jelena Stanojevi\'c, \tt jelenas@ekof.bg.ac.rs\\ \it University of Belgrade, Faculty of Economics and Business}

\date{}

\begin{document}

\maketitle

%\large

\begin{abstract}
\normalsize  In this paper we consider systems which consist of binary components with known reliabilities. We discuss their algebraic properties and define the  corresponding algebraic structure, which we call the reliability algebra. We prove that the reliability algebra is a Boolean algebra. The reliability algebra seems an appropriate context for defining fuzziness measure i.e. membership function with reliability values.
\end{abstract}


AMS 2020: 03B52 03G05 90B25 62N05 60K10

%\end{frontmatter}

%\linenumbers

%Abbreviated title: Boolean reliability algebra

%\newpage

%Corresponding author: Jelena Stanojevi\' c, jelenas@ekof.bg.ac.rs

%\newpage

\section{Introduction}

The framework of random events algebra, as usually used in probability theory, is not always good enough for modeling uncertain, unreliable, vague or imprecise real life phenomena. After explosion of Zadeh's fuzzy systems theory based on the notion of fuzzy set (see \cite{Zd}), and Zadeh's fuzzy logic concept (see \cite{Zd2}), later, between other significant alternatives made for dealing with uncertainties, we mention Shafer's evidence theory (see \cite{Sh} or \cite{KM}), Pawlak's rough set theory (see \cite{Pw} or \cite{ZWX}) and Molodtsov's soft set theory (see \cite{Mol1} or \cite{Mol2}).

Each Heyting--valued function and, consequently, each Boolean--valued function, presents potentially a possibility for fuzzification of any structure of crisp sets (see \cite{Boricic2}). On the other side, reliability of a system, as characteristic property, by its nature, can be considered a measure of vagueness, uncertainty or even of logical inconsistency of the system. In this paper we prove that reliabilities, defined in an appropriate way, have Boolean structure, qualifying them for fuzziness measure i.e. for values of membership function in a fuzzy system. In particular, herewith we prepare a ground for formalism enabling to express 'reliability of a statement' and 'reliability of a proof'. Namely, when we prove that a system consisting of reliabilities $r$ has a Boolean structure, then we sense that it is possible to formalize a sentence '$r$ is reliability of a statement $A$', denoted by $A^r$, enabling to work with more complex entities such as $A^r,B^s\vdash C^t$, meaning that 'conclusion $C$, with reliability $t$, can be derived from hypotheses $A$ and $B$, with reliabilities $r$ and $s$, respectively (see \cite{Boricic}, \cite{Boricic3}). The fact that a system of reliabilities presents a Boolean algebra, which is proved in this paper, presents the first and basic step in building a structure making possible to define reliability $r$ as fuzziness measure , value of membership function, of a statement $A$, formally working with expression of the form $A^r$.

The reliability of a system, possibly consisting of many components,  is the probability that a system will function. It is a significant notion in statistics and information theory whenever  it is necessary to define the sureness of an information or of a system. The reliability function is then usually introduced by means of probability distribution function of the corresponding random variable and the problem of reliability of a complex system is considered via  probabilistic characteristics of the consisting elements (see \cite{Barlow}, \cite{Gertsbakh}, \cite{Ross}). The same function appears also in models of actuarial mathematics, under the name of the survival function (see \cite{Gerber}).

Also, it is important to say that one applicable part of reliability mathematics are for example, shock models. In the so far literature two basic types of that models are studied: cumulative shock models and extreme shock models, theirs details are out of scope of this paper. But it is worth to point out that shock magnitude is measurable and the system life time of the above models is defined in its terms. New approach, with $\delta$-shock model was proposed in the 1999. (see \cite{Li} and their previous work) and it is developed until now (see \cite{Ming and Li}). In that model a present parameter $\delta$ determine critical interval $C(\delta)$, and the system fails when the interval of two successive shocks enter into that interval $C(\delta)$. This $\delta$-shock models finds their applications in many scientific fields like: earthquake modeling, insurance mathematics, electrical systems, inventory theory and many others reliability areas. These above facts can be a good reason for further and deeper studying of the topic of this paper, in particular because of its importance and applications from different point of view.

Specially, the correspondence between the basic propositional connectives and binary block diagram connectives is well known, e.g.  conjunction corresponds to combining of two blocks in series (or cascade) structure, disjunction to combining of  two blocks in parallel, and implication to combining two blocks in a feedback control system (see \cite{Distefano}). In this paper we consider this relationship with the respect to the reliabilities of the binary block diagram components (not just with the respect to the information whether they are functioning or not). The central point is to justify that the reliability algebra defined here has a Boolean structure (see \cite{Givant}). This gives the formal background for reliability calculations, due to the fact that the Boolean techniques are usually used for that purpose.

%On the other hand there is an one--to--one correspondence between systems of diagrams and systems of reliabilities helping us to understand better functioning the complex systems consisting of mutually connected elements with known reliabilities and to analyze them.

An immediate motive to study  the basic algebraic properties of reliabilities rise from our investigation of a reliability logic (see \cite{Boricic}, \cite{Boricic3}) where a more precise definition of a reliability algebra is required. Our present result also gives the needed verification of that  subject.

\section{Algebra of block diagrams is a Boolean algebra}

A {\it Boolean algebra} is an algebraic structure  ${\cal B}=\langle B,\i,\ili,',1,0\rangle$, where $B$ is a non--empty set with at least two distinct elements 0 and 1, two binary operations $\i$ and $\ili$ and one unary operation $'$, such that $\i$ and $\ili$ satisfy the commutative and the associative laws, as well as the distributive law of $\i$ with respect to $\ili$, and vice versa, together with the neutral elements laws:
$x\i0=x$, $x\ili 1=x$, and  the complement laws (or the  inverse laws): $x\ili x'   =1$ and $ x\i x'=0$, where $x'$ is called the complement of $x$ (see \cite{Givant}).

Boolean terms are defined inductively, as usual. Two Boolean terms $t_1$ and $t_2$ are {\it equal}, denoted by $t_1=t_2$, iff we can transform one of them into the other one, using only the axioms of the Boolean algebra and basic properties of equality.

The  simplest Boolean algebra is a  two--element Boolean algebra \linebreak  ${\cal B}_2=\langle \{0,1\},\i,\ili,',1,0\rangle$. Variables of this algebra can take only the values 0 or 1. This dichotomy applies also to  Boolean terms, such that  whether a Boolean term  takes the value 0 or 1 is determined completely by the values of its variables. It is worth mentioning that when the elements 0 and 1 of a two--element Boolean algebra are the whole numbers, then the operations of the Boolean algebra can be understood as: $x\i y=\min(x,y)$, $x\ili y=\max(x,y)$ and $x'=1-x$. Our two--element Boolean algebra is then ${\bf 2}=\langle \{0,1\},\min,\max,1-\phantom{x},1,0\rangle$.

One well--known example of ${\cal B}_2$ algebra is the algebra of binary block diagrams. A binary  block diagram (or simply, a diagram) is a structure  whose every component is either in the functioning or in the failure state and which itself can also be either functioning or has failed. We use ${\bf 1}$ to denote functioning diagram and ${\bf 0}$ to denote a diagram which has failed. In order to define an algebra of diagrams we  choose the following three operations, two of them are binary (we choose them among $2^{2^2}$ binary operations which can be defined over the set $\{{\bf 1,0}\}$), denoted by $\di$ and $\dili$, and one of them is unary (this is one of $2^{2^1}$ unary operations which can be defined over the set $\{{\bf 1,0}\}$), denoted by $\ol{\phantom{x}}$, defined as follows:$$\begin{array}{c|cc}\di&{\bf 1}&{\bf 0}\\ \hline{\bf 1}&{\bf 1}&{\bf 0}\\{\bf 0}&{\bf 0}&{\bf 0}\end{array}\qquad\begin{array}{c|cc}\dili&{\bf 1}&{\bf 0}\\ \hline{\bf 1}&{\bf 1}&{\bf 1}\\{\bf 0}&{\bf 1}&{\bf 0}\end{array}\qquad\begin{array}{c|c}&\ol{\phantom{A}}\\ \hline{\bf 1}&{\bf 0}\\{\bf 0}&{\bf 1}\end{array}$$which we call {\it series, parallel} and {\it complement}, respectively.

Then  an algebra of diagrams is defined as follows.\medskip

\noindent{\bf Definition.} An {\it algebra of diagrams} is an algebraic structure $${\cal D}=\langle\{{\bf 0}, {\bf 1}\},\di,\dili,\ol{\phantom{x}},{\bf 1}, {\bf 0}\rangle.$$

We use $A,B,C,\dots$, with or without subscripts, to denote binary components and we define diagrams, inductively, as follows:\medskip

\noindent{\bf Definition.} (i) Binary  components $A_1,\dots,A_n$ are {\it diagrams}. We shall call them {\it elementary diagrams}.

(ii) If $D$, $D_1$ and $D_2$ are diagrams then $(D_1\di D_2)$, $(D_1\dili D_2)$ and $\ol{D}$ are also {\it diagrams}.

(iii) {\it Diagrams} can be obtained by applying only (i) and (ii), finitely many times.\medskip

We shall use $D$, with or without subscripts, to denote diagrams and we shall omit, as usual, the outer pair of parentheses.

By the definition or our connectives,  it is clear that the  diagram $D_1\di D_2$ is in the functioning state iff both $D_1$ and $D_2$  are  functioning, the diagram $D_1\dili D_2$ is in the functioning state iff at least one of $D_1$ or $D_2$  is  functioning, and, finally, the diagram $\ol{D}$  is in the functioning state iff $D$ has failed.

Moreover, our connective $\di$ is associative,  in the following sense.  If $D_1,D_2,D_3$ are diagrams, then the diagram $(D_1\di D_2)\di D_3$ is in the functioning state iff all of $D_1,D_2,D_3$ are functioning; similarly, the diagram $D_1\di (D_2\di D_3)$ is in the functioning state iff all of $D_1,D_2,D_3$  are  functioning, therefore  $(D_1\di D_2)\di D_3$ is functioning  iff $D_1\di (D_2\di D_3)$ is functioning, and we write this fact as:$$(D_1\di D_2)\di D_3=D_1\di(D_2\di D_3).$$

Analogously we can verify that $\di$, $\dili$, $\ol{\phantom{x}}$, ${\bf 0}$ and ${\bf 1}$ satisfy all axioms of Boolean algebra, so we have:\medskip

\noindent{\bf Theorem.} {\it ${\cal D}$ is a Boolean algebra.}\medskip

Consequently, diagrams can be understood as Boolean terms, and it is natural to define the equality between two diagrams as follows.\medskip

\noindent{\bf Definition.} Two diagrams $D_i$ and $D_j$ are {\it equal}, denoted by $D_i=D_j$, iff they are equal as Boolean terms.
\medskip

In order to determine whether an arbitrary  diagram $D$ is in the functioning or in the failure state, usually  the {\it structure function} $\Phi$ is defined, as follows:$$\begin{array}{c}\Phi({\bf 0})=0,\qquad \Phi({\bf 1})=1,\qquad  \Phi(\ol{D})=1-\Phi(D),\\\Phi(D_1\di D_2)=\min(\Phi(D_1),\Phi(D_2)),\\ \Phi(D_1\dili  D_2)=\max(\Phi(D_1),\Phi(D_2)).\end{array}$$

Then, $D$ is in the functioning state if $\Phi(D)=1$, and it is in the failure state, otherwise, i.e. when $\Phi(D)=0$.

Immediately, we have:\medskip

\noindent{\bf Theorem.} {\it The structure function $\Phi$ is an isomorphism between the algebras ${\cal D}$ and ${\bf 2}$.}\medskip

This is a kind of a 'semantical' point of view. Namely, any diagram is interpreted as either functioning or not, i.e. as a single value in the set $\{0,1\}$. However, in the sequel we shall be interested in another interpretation of binary diagrams. Namely, we shall interpret diagrams in a set of  {\it reliability terms} and that will take us closer to a kind of a 'syntactical' point of view.

\section{Reliability algebra is a Boolean algebra}

It is well--known that the  {\it reliability of a diagram $D$},  denoted by $r(D)$, is  the probability that the diagram $D$ is in the functioning state, i.e. that $r(D)$ is a real number between 0 and 1. However,  instead over a set of real numbers, we shall  define a reliability algebra of diagrams over  a  set of {\it  reliability terms}. They should be understood as terms built--up upon some finite set of letters, which will be called {\it reliability constants}. \medskip

We shall consider only diagrams which are built--up upon a set of $n\geq 1$ mutually nonequal elementary diagrams. Accordingly, we introduce the following notions:\medskip

\noindent{\bf Definition.} The set ${\bf A_n}=\{A_1,\dots,A_{n}\}$  will be called the {\it generating set} when $A_1,\dots,A_{n}$ are mutually nonequal  elementary diagrams.  We say that a diagram $D$ {\it is built--up upon} ${\bf A_n}$ iff  every component of  $D$ is in ${\bf A_n}$.\medskip

It is clear that there are exactly $2^{2^n}$ mutually nonequal diagrams, built--up upon ${\bf A_n}$. Recall that, this is the same as in the propositional logic, where the number $2^{2^n}$ is exactly the number of mutually nonequivalent  formulas built--up upon the set of $n$ propositional letters. Moreover, if ${\bf For_n}$ is the set of all mutually nonequivalent formulas      built--up upon a set of $n$ propositional letters and  ${\bf Dia_n}$ is the set of all mutually nonequal diagrams   built--up upon a generating set of $n$ elementary diagrams, then we have:\medskip

\noindent{\bf Theorem.} {\it The algebraic structure $For=\langle{\bf For_n}, \wedge,\vee,\neg,\top,\bot\rangle$  is isomorphic to the algebraic structure $Dia=\langle{\bf Dia_n}, \di,\dili,\overline{\phantom{A}},{\bf 1},{\bf 0}\rangle$.}\medskip

Let us note that the proposition 'a diagram is   built--up upon a generating set' is  equivalent to the condition that a diagram is built--up upon  {\it independent components}, which is, as is well--known, the crucial condition for expressing the reliability of the system as the function of the component reliabilities (see \cite{Ross}).

Now, a {\it reliability algebra} can be defined as follows:\medskip

\noindent {\bf Definition.} Let ${\bf A_n}$ be a generating set and let $R=\{r_1,\dots,r_n\}$ be such that $r_i$ is a {\it reliability constant} assigned to an elementary diagram $A_i$, for every $1\leq i\leq n$.  Then ${\cal R}=\langle{\bf R},\circ,\ddagger,\phantom{x}^{-1},i,o\rangle$ is called a {\it reliability algebra} of diagrams built--up upon ${\bf A_n}$, where  ${\bf R}$ is  the algebraic closure of the set $R$ with respect to binary operations $\circ$ and $\ddagger$ and a unary operation $\phantom{x}^{-1}$, where $\circ$ and $\ddagger$ satisfy  commutative and  associative laws, as well as a distributive law of $\circ$ with respect to $\ddagger$, and a distributive law of $\ddagger$ with respect to $\circ$ and where $i$ and $o$ are such that, for every $x\in {\bf R}$ it holds that  $x\ddagger x^{-1} =i$ and $ x\circ x^{-1}=o$ and $i$ and $o$ are neutrals for $\circ$ and  $\ddagger$, respectively,  i.e. they satisfy the neutral element laws: $x\ddagger o=x$ and $x\circ i=x$, for every $x\in{\bf R}$.\medskip

As an immediate consequence we have:\medskip

\noindent{\bf Theorem.} {\it The reliability algebra of diagrams built--up upon ${\bf A_n}$ is a Boolean algebra.}\medskip

It is also clear that the cardinality of $\bf R$ is exactly the same as the cardinality of the set of diagrams built--up upon $\bf A_n$. Thus, to each of $2^{2^n}$ mutually nonequal diagrams built--up upon $\bf A_n$, a unique term, called the {\it reliability term} is assigned.

Another, practical question is how can we effectively calculate the reliability of an arbitrary diagram $D$, built--up upon $\bf A_n$? The question is in fact how to identify the diagram equal to $D$, whose reliability term is in $\bf R$. But, this has been studied in the literature (see \cite{Bennetts}, \cite{Gupta}) and  is not  the subject of this paper.\medskip

\section{Concluding remarks}

An approach to modeling reality with undetermined, vague and stochastic elements may be founded on reliability as its basic notion. Those models include statistical, technical, logical etc. treatments of complex systems built up upon elements with given estimated times of their functioning life. Reliability function appears as a basic notion. A traditional probabilistic reliability analysis of systems is usually a part of textbooks in probability theory (see \cite{Ross}). Our interest in an abstract algebraic structure whose  set contains  elements with given reliabilities, is inspired by the problem how to define the reliability of a complex sentence consisting of atoms with known reliabilities (see \cite{Boricic}, \cite{Boricic3}). Namely, the structure of any sentence, made by means of connectives, points to an analogy with a diagram. This point of view leads us to consider sentences as diagrams, as well as sentence's reliability as diagram's reliability. Our main conclusion is that the space, or the algebra, consisting of reliabilities as its basic elements, is structured as a Boolean algebra (see \cite{Givant}). This fact will be of great interest in further investigations of reliability logic enabling one to express and calculate the reliability of a sentence by working formally with propositions of the form $A^r$ with the intended meaning that '$r$ is the reliability of the sentence $A$' in a framework of a logical reasoning system.

\end{document}